# Analyzing Bank Account Information of Nominees and Scammers

Patsita Sirawongphatsara

Phisit Pornpongtechavanich

Pakkasit Sriamorntrakul

Therdpong Daengsi

**Abstract:** Nowadays, people heavily rely on the Internet for various activities, such as e-commerce (e.g., online shopping) and online banking. While online transactions are practical, they also provide scammers with a new way to exploit unsuspecting individuals. This study and investigation utilized data from ChaladOhn, a website designed and developed by academics and policemen. The data covered the period from February 2022 to January 2023. After analyzing and investigating, the results reveal that the total losses amounted to over 3,100 million Thai Baht, with each case incurring losses of less than 10 million. Furthermore, the investigation discovered the involvement of the top two banks in the market, KB*** and BB*, in the fraud. These banks accounted for: 1) 28.2% and 16.0% of the total number of scam accounts, 2) 25.6% and 20.5% of the total transactions, and 3) 35.7% and 14.9% of the total losses from the victims as recorded in the database, respectively. Considering the anticipated deterioration of this issue, it is crucial to inform regulators and relevant organizations about the investigation's findings. This will enable the development, suggestion, and implementation of an efficient solution to address the rapidly increasing number of online scam cases.

# 1. Introduction and Related Work

E-commerce is already a common practice, particularly in Asia, which is the world's largest retail e-commerce market and is predicted to generate $2,055 billion in revenue this year (see Fig. 1). [1]. The COVID-19 pandemic that will hit between 2020 and 2022 is one of the factors accelerating the expansion of e-commerce. People had to maintain their distance, which led to an increase in e-commerce, particularly online purchasing. But since the number of websites has increased, be aware of phishing attacks because by carelessly clicking on these links, harmful URLs and user data such login information and credit card numbers can be obtained [2]. However, one issue with e-commerce is the prevalence of online frauds. It has been shown that scammers use a variety of cunning methods to obtain the personal data of their victims, including:

- Create a false identity as a bank or government representative, then phone the victim and demand personal information.
- Call or solicit victims to click on phony links that contain malware or other harmful software
- Create listings on an e-commerce site, but fail to fulfill the victims' orders for the products.
- Promotion of an online casino
- Request investments from the victims in phony securities that offer extremely high returns.
- Romance scams

Once the scammers have the victims' personal information, they use it to gain unauthorized access to their bank accounts and transfer money to money mules' accounts, also known as Ban Shee Ma (literally, "horse accounts" in Thai), which are individuals who move or transfer illegally acquired money on behalf of another. [3]

The first layer mules then pass the victims' money to the second and upper tiers mules, as indicated in Fig. 1 [4], respectively. The money will then be given to the Boss who has an electronic wallet and/or a cryptocurrency wallet. It is obvious that practically all of the accounts used by scammers are those of money mules that are dispersed throughout numerous Thai banks and financial organizations. This study used data from the ChaladOhn [5] database, a website created in partnership with the Thai Metropolitan Police Bureau to protect Thais from online scams, to examine the characteristics of the spread of money mule accounts.

This study's main contribution is the disclosure of Thailand's top five banks with the most accounts belonging to money mules and scammers as well as the loss amounts from victimization.

This article is the expanded version of [6]. For purposes of structure, Section 2 presents the background data and literature review following the presentation of this section. The methodology is then demonstrated in Section 3. In Sections 4 and 5, they provide interesting figures and discussions to describe the findings and analyses, respectively. Finally, as in Section 6, the conclusion is stated.

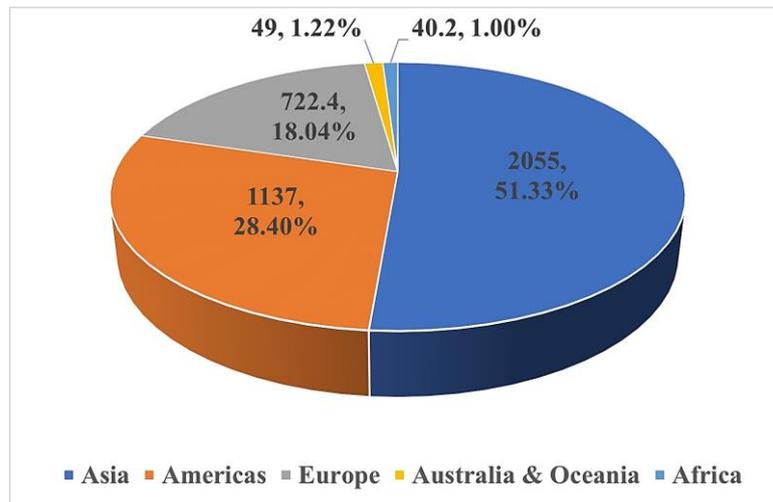

Fig.1. By region, global retail e-commerce revenue in billion USD in 2023 [1]

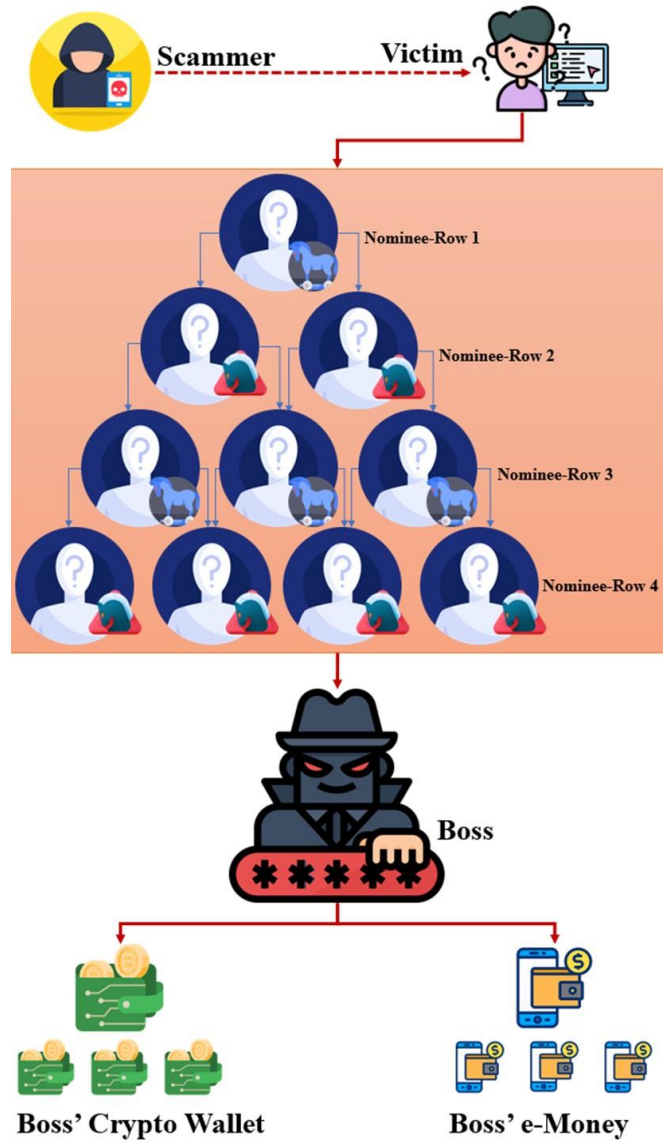

Fig.2. Typical money flow in well-organized online frauds is described [4]

## 2. Background Information and Review of Previous Studies

### 2.1. Online Scams

As mentioned in [7], scammers use a variety of methods, including the following, to commit fraud or steal money from victims who have bank accounts, mobile banking apps, or internet banking apps:
1) Fraud SMS: In this method, scammers send phony SMS messages to their victims under the guise of a financial institution, a reputable company, or a well-known phone number. However, some suspicions are obvious because the text messages frequently induce fear and apprehension in the recipients or excitement and contentment as if they will receive some privilege or award, and they press the recipients to act quickly before their accounts expire. They may also tempt recipients to click an attached link to enter personal information in exchange for a free gift. Complete information, such as a person's ID, bank account number, credit card number, birthdate, ATM code, and/or password, will be requested in fake mail. Additionally, users will reach fraudulent websites if they click the link included in the SMS.
2) Fake website: Users can first examine the URL of the financial institution to gain confidence and prevent online attacks. Some URLs can be used to determine the nation where a website is registered; for instance, https://www.scb.co.th indicates that the website was created in Thailand. Users can verify the website information for URLs with suffixes like.com, .net, or other without country suffix at https://www.whois.com/whois/, which reveals where that website is registered and who the owner is. When users click it, if it's a phony, they'll be prompted to enter extensive personal data. Users should not believe it is unusual as a result. Typically, the bank simply asks for the most basic data, like name, last name, email, and phone number.

3) Fake LINE: In Thailand, banks or other businesses sometimes use official LINE profiles. As a result, scammers use this channel of contact to obtain personal data from bank customers in order to gain access to their bank accounts. Typically, a false LINE account will start by adding and welcoming people. Users can see "Add friends" at the top, indicating that they have not yet become friends with this account. In contrast, a real LINE account requires the user to manually add friends. Additionally, since most bank employees communicate with consumers using chatbots rather than personally, the LINE account is unable to initiate the first conversation with any customer. Naturally, detailed personal information may be requested during the chat, or the user may be prompted to visit a phony link, website, or application in order to hijack their bank account. The victim's real friend's LINE account can also be hacked, or a new line account can be created using the real friend's profile, for example, before requesting a loan with a false justification.
4) Facebook fraud: in this instance, scammers create a phony Facebook page for a bank or other financial institution. Because it frequently resembles a legitimate Facebook page and the submitted contents also appear authentic, many bank customers may not realize it is a scam if they are not paying close attention. If they look closely, though, they might be able to spot suspicious items. The fake Facebook may utilize names that are nearly identical to the genuine Facebook but contain specific indicators (such as commas, full stops, or other uncommon characters) that could confuse the bank customer. Moreover, the fake bank Facebook page might only have a few hundred fans, in contrast to the millions of fans on the actual page. Additionally, every post on the genuine page typically receives hundreds or thousands of likes, as opposed to every post on the phony page receiving none or a small number of likes.
5) Spam email: In general, Thai banks don't email clients to request confidential or sensitive information (such as a customer ID or citizen ID, PIN code, or essential information over email). There is no bank policy to contact a customer directly from a bank officer to ask for secret personal information or to do things that may be risky to be hacked, [8] so customers should be aware of emails that ask them to disclose sensitive or personal information, update information, change passwords, or click external links.
6) Scam App: It is currently taking off and becoming popular among scammers. They use LINE or fraudulent SMS to deliver links to malicious applications that can be downloaded. They frequently use numerous offers that are scams to trick their victims, such as loans with extremely cheap interest rates. The victim will be taken directly to the program if they choose to download rather than going through the Play Store, App Store, or Huawei App Gallery. There is no application information provided for this type of application notice, such as the number of reviewers, the number of downloads, rating information, or file size.

Of course, users of those applications and banks must be cautious, aware, and observant to avoid becoming victims, regardless of the methods used by fraudsters or scammers.

### 2.2. Predicted Worldwide Cybercrime Loss

According to [1][9], the estimated global loss from cybercrime in 2025 would be over 17.7 trillion USD (as shown in Fig. 2), which is a significant increase from the estimated loss of 10.5 trillion USD in 2025 given in the Cybercrime Magazine [10]. Despite the fact that the two anticipated numbers differ, they show that cybercrime is a major global problem. The estimated cost from cybercrime was based on previous cybercrime data, which also included statistics on organized crime actions by hacking groups. The damage or destruction of data, lost productivity, theft of financial, intellectual, and personal data, fraud, embezzlement, and theft of money are just a few examples of the losses that can result from cybercrime [10]. Most cyberattacks target financial institutions. This reduces the number of intermediaries an attacker needs to hit to reach the target. [11] Other losses might include forensic investigation, reputational harm, and post-attack business disruption.

### 2.3. Overview of Online Scams in Thailand

Online fraud has recently become a significant problem in Thailand as a result of the proliferation of fraud tactics used by scammers, such as false loan applications, phony call centers, remote access malware for smartphones, and misleading text messages. According to online police reports from March to December 2022, fraudulent listings accounted for 32.6% of all online scams, followed by misleading online employment, false online loans, online investment fraud, and scam call centers (see Fig. 3). This is in keeping with what was previously stated in [12].

The Bank of Thailand (BOT) claimed that it continually develops and implements procedures to stop these internet scams. BOT has also worked together with the relevant authorities to put these safeguards in place to combat cybercrime. Almost twenty banks are among the financial institutions that the bank regulator has ordered to regularly update their systems to combat cybercrime and improve joint operation with the appropriate parties to prevent such crimes.

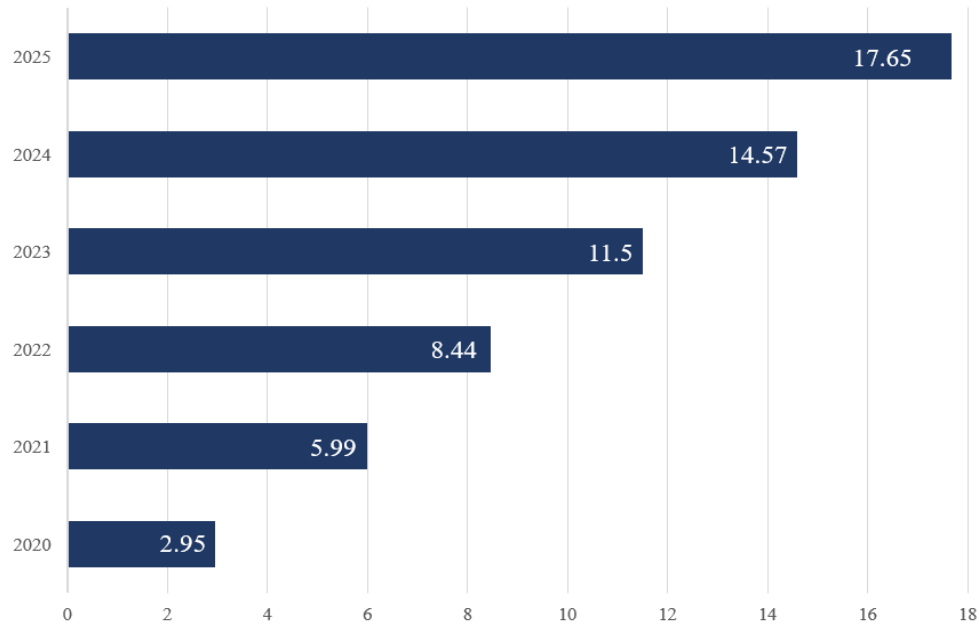

Fig.3. Top 5 internet complaints to the police, adapted from [12]

Table 1. Statistic of online police complaints [12]

|  | Online sales | Online employment | Online faulty loans | Online investment | Call center |
|---|---|---|---|---|---|
| No. of cases | 53,080 | 22,781 | 19,349 | 14,313 | 13,178 |
| Percentage | 32.6% | 14.0% | 11.9% | 8.8% | 8.1% |
| Loss value (million THB) | 750 | 2,562 | 819 | 6,977 | 2,620 |

*2.4 Previous research works*

Following a study of prior studies, numerous articles with strong arguments and insightful debates about online fraud were discovered. They fit the description given in Tale 2:

Table 2. Previous works

| Authors | Findings |
|---|---|
| Kale et al. [13] | They suggested a method that can identify phishing phone calls by listening to the discussion between the victim and the fraudster. They conducted an intent analysis of call transcripts using a variety of machine-learning approaches. According to their research, CNN-based models have a peak accuracy of 97.21%. |
| Datta et al. [14] | They did a thorough analysis of the numerous frauds that occurred online or through mobile banking applications and services, concentrating on the rise in online fraud instances involving the banking sector. They came to the conclusion that education campaigns are required to stop or minimize online fraud. |
| Singh et al. [15] | They investigated fraud (via the Punjab National Bank case) in an effort to pinpoint the contributing reasons. They suggested that all financial institutions watch every employee closely and that the procedures for markers and checkers be enhanced. |
| Sharma et al. [16] | By adopting a more dependable and effective login procedure using a variety of approaches and applications (e.g., third-party apps, IP address, region (location) to log in, API, OTP), they recommended a design change to the current banking system. This is done to lessen the likelihood that customers' sides will experience security breaches. |
| Siddiqui [17] | Biometric authentication is one of the methods that can be used to increase security, according to Siddiqui. In addition to PIN authentication and verifications, biometric authentication methods like fingerprints, iris scans, palm scans, and even voice recognition can be used to prevent ATM fraud and enhance the security of other financial transactions. |

| Singh et al. [18] | They suggested locality-based machine learning be used in conjunction with domestic legislation that complies with international norms. This three-layer security method, which combines domestic law, international standards, and machine learning, could address the security and privacy concerns brought on by financial fraud involving social networking sites (SNS) in order to decrease financial fraud. |
|---|---|
| Rawandale et al [19] | They performed the study and showed how clients of online banking services can suffer as a result of financial companies' violation of precautionary norms. As a result, it's important to safeguard clients from fraud or exploitation by passing the proper legislation and upholding the law. |
| Roy et al. [20] | To spot Master Card fraud, they deployed three machine learning algorithms. They indicated that the bank may use their techniques, including Isolation Forest, to identify credit card scams. However, additional steps should be taken. |
| Kameel et al. [21] | They demonstrated how the legislator in the UAE offers decentralized protective implementation across various related laws, which is a significant step toward innovative legislative solutions to offer state-of-the-art protection for UAE consumers during online transactions against false deceptive advertisements. |
| Nandhini et al. [22] | The majority of social network users are unaware of the various security threats and associated risks on the social network, according to a study they conducted on social networks. They found that it is the main method scammers use to obtain sensitive information about their victims for use in various fraudulent activities. |
| Yoshida et al. [23] | They offered an innovative approach that focuses on exploiting the traits of economic and corporate crimes, such as online shopping fraud committed by dishonest customers. Short-term apartment rentals and the use of deferred payment schemes were found to be two important indicators in Japan. They also discovered that scammers frequently purchase products that can be quickly sold for cash. |
| Daengsi et al. [24] | They performed an analysis using the information from the ChaladOhn website's database [3]. They discovered that the victims' average age ranged from 20 to 39. Therefore, it is crucial to spread knowledge and provide education about technical literacy to young Thais. |
| Al-Ali & Al-Nemrat [25] | They employed a method of online surveying pertaining to cybercrime and cybervictimization. Identity fraud, cyberharassment, and cyberattack are the main trends of cybervictimization in the UAE, according to the data. Strong correlations between online behavior, online time, and cybervictimization have been shown by the evidence. The chance of becoming a cybervictim was found to be related to technological guardianship, online behavior and usage, computer skill, time spent online, region of residence, and gender, according to a logistic regression study. |
| Punkamol & Marukatat [26] | They carried out research in relation to a system to identify account cloning in online social networks. Twitter Crawler, Attribute Extractor, and Cloning Detector are the three components that make up the framework. As a case study, Twitter was used. The framework has an average accuracy of 80% in identifying whether the posts were phony or real. While decision trees produced the best categorization results, it was discovered. |
| Palad et al. [27] | The Weka text mining technology is used in this work to examine a dataset of online scams based on Filipino terms. J48 Decision Tree, Naive Bayes, and Sequential Minimal Optimization are used to process the dataset. J48 outperforms the competition with the highest accuracy and lowest error rate, followed by Naive Bayes and SMO, according to the data. J48 is also preferred for its practicality and user-friendliness in cybercrime investigations. |
| Anupriya et al. [28] | They demonstrated the effectiveness of supervised learning in neural network computations, resulting in increased accuracy with a wide scope for misrepresentation and a lower rate of false alarms. Well-trained ANNs show their potential in this area by mimicking the way neurons work in the human brain. Machine learning is used for false detection, utilizing user transaction records to examine e-commerce-related behavior patterns. |
| Fkih & Al-Turaif [29] | They have done research on designing a twitter threat detection model using semantic networks developed called DetThr, the model was developed to detect threatening and fraudulent tweet messages. it can be used to reduce crime that will occur in the future. |

These works demonstrate that the majority of earlier works, particularly [14–20], concentrated on the financial sector because banks are the main target of fraudsters and scammers when they try to steal money. As a result, it would be helpful to evaluate the data from ChaladOhn [5] to identify problems and weaknesses related to the bank sectors.

## 3. Proposed Methods

The data set for this investigation was exported from the database of ChaladOhn website that was a collaborative project between the Faculty of Engineering at Rajamangala University of Technology Phra Nakhon (RMUTP) and the Investigation Division of Metropolitan Police Division 8, funded by the Broadcasting and Telecommunications Research and Development Fund for Public Interest [5][30]. The information spans the months of February 2022 and January 2023. The information was gathered from actual online fraud victims and perpetrators. However, only information related to online scammers was used in this study.

There are more than 87,500 records in the data. Any duplicate information from the con artist and/or money mules with the identical names and bank account numbers was taken into consideration and eliminated during the preparation process. It was also discovered that the instances with losses of more than one million Thai Baht (THB) made up a very little part of the total loss value arising from the cases with losses of less than one million THB each.

## 4. Results and Analysis

This study used data from the ChaladOhn website's database with losses ranging from 100 THB to 10,000,000 THB. There were total losses of 4,685,252,172 THB, total transactions of 87,943, and total accounts of 61,805 scammers or money mules. The following three major points are highlighted:

### 4.1 Statistics Based on Loss Values

Following the deletion of duplicate bank accounts, situations with a wide range of loss values were divided into the following five categories:
  (a) 100-1,000 THB per transaction or called Category A
  (b) 1,001-10,000 THB per transaction or called Category B
  (c) 10,001-100,000 THB per transaction or called Category C
  (d) 100,001-1,000,000 THB per transaction or called Category D
  (e) 1,000,001-10,000,000 THB per transaction or called Category E

The data were then utilized to produce pie charts as displayed in Fig. 4(a) through (e).

The following details are derived from the results associated with the loss values shown in Fig. 4:
- As shown in Fig. 4(a), KB*** is ranked first in Category A with 26.1% of the loss values between 100 and 1,000 THB per transaction, followed by BB*, SC*, KT*, and GS*, which are ranked second through fifth, with respective loss values of 20.0%, 13.1%, 11.2%, and 8.3%.

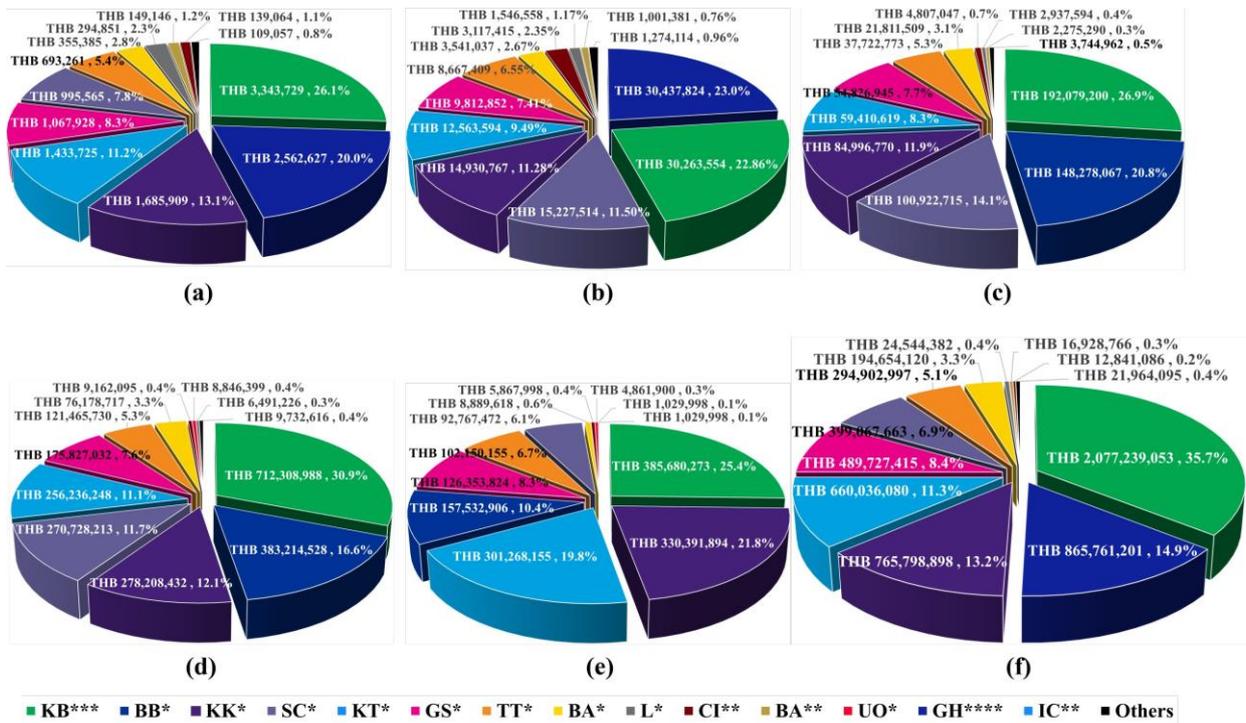

Fig.4. The statistics based on loss values (THB) per transactions by bank (a) 100-1,000 (b) 1,001-10,000 (c) 10,001-100,000 (d) 100,001-1,000,000 (e) 1,000,001-10,000,000 (f) Overall

- The position changed when Category B, the loss values between 1,001 and 10,000 THB per transaction, were considered. In Fig. 4(b), it can be seen that BB* is ranked first with 23,0% of the loss values, slightly higher than KB***, who is in second place with 22,9%. KK*, SC*, and KT* are in third through fifth place, respectively, with 11,5%, 11,3%, and 9,5% of the loss values.
- On the other hand, for Category C, the loss values between 10,000 and 100,000 THB per transaction, as shown in Fig. 4(c), KB*** reclaimed the top spot with 26.9% of the loss values, followed by BB*, KK*, SC*, and KT* with 20.8%, 14.1%, 11.9%, and 8.3%, respectively.
- It is comparable to Fig. 4(a) for Fig. 4(d). Among transactions involving 100,000 THB or more, or Category D, KB*** is rated first with 30.9% of the loss values, followed by BB*, SC*, KK*, and KT*, which are ranked second through fifth, with 16.6%, 12.1%, 11.7%, and 11.1% of the loss values, respectively.
- Contrary to other data, Fig. 4(e) shows that for Category E, the loss values between 100,000 and 1,000,000 THB per transaction, KB*** still holds the top spot with 25.4% of the loss values, but SC*, KT*, BB*, and GS* are surprisingly in the second to fifth positions with 21.8%, 19.8%, 10.4%, and 8.3% of the loss values, respectively.
- Lastly, the overall figure in Figure 4(f) may be seen to be identical to Figure 4(a). KB*** is placed highest with 35.7% of the loss values, followed by BB*, SC*, KT*, and GS* in that order, with respective loss values of 14.9%, 13.2%, 11.3%, and 8.4%.

*4.2 Statistics Based on Number of Transactions*

The information can be summarized as follows based on the outcomes connected to the transaction numbers shown in Fig. 5:

• From Fig. 5(a), it can be seen that KB*** is placed first with 27.6% of the transactions linked with losses of between 100 and 1,000 THB per transaction, or Category A, while BB*, SC*, KT*, and GS* are ranked as the second to fifth position with 18.2%, 14.1%, 11.9%, and 8.6%, respectively.

• When transactions with loss values between 1,001 and 10,000 THB per transaction, or Category B, were considered, it can be seen in Fig. 5(b) that KB*** is ranked first with 22.7% of the transactions with loss values, slightly higher than BB*, who is in second place with 22.3%. SC*, KK*, and KT* are in third through fifth places, respectively, with 11.3%, 10.9%, and 9.9% of the loss values.

• However, as shown in Fig. 5(c), for transactions associated with Category C, the loss values range from 10,001 to 100,000 THB per transaction. KB*** takes the top spot with 25.8% of the transactions associated with the loss values, followed by BB*, KK*, SC*, and KT* with 21.8%, 14.1%, 11.8%, and 8.1%, respectively.

• The situation is comparable to Fig. 5(c) for Fig. 5(d). KB*** is ranked first with 29.8% of the transactions belonging to Category D, or the loss values between 100,001 and 1,000,000 THB per transaction. BB*, KK*, SC*, and KT* are ranked second through fifth, with 17.3%, 13.2%, 11.8%, and 9.7% of the loss values, respectively.

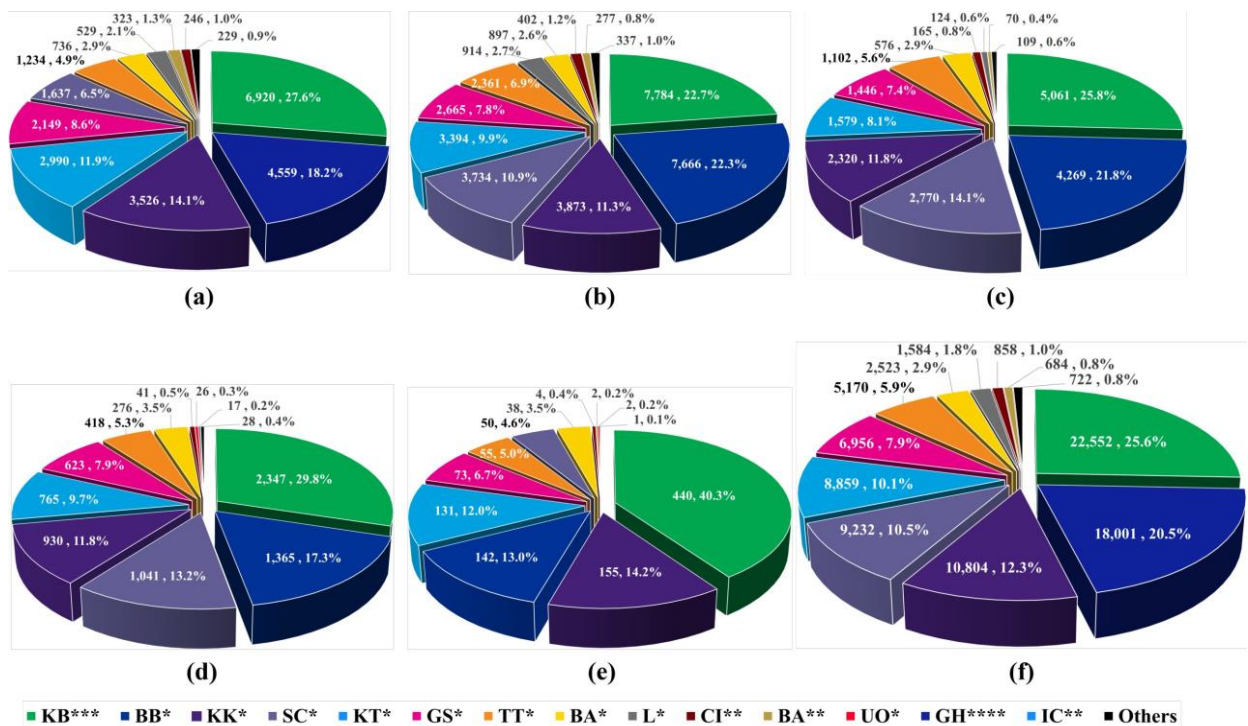

Fig.5. The statistics based on numbers of transaction (THB per transactions) by bank (a) 100-1,000 (b) 1,001-10,000 (c) 10,001-100,000 (d) 100,001-1,000,000 (e) 1,000,001-10,000,000 (f) Overall

- According to Fig. 5(e), the first place is still held by KB*** with 40.3% of the loss values, but surprisingly, SC*, BB*, KT*, and GS* are in the second through fifth places with 14.2%, 13.0%, 12.0%, and 6.7% of the loss values, respectively. For Category E, the loss values between 100,001 and one million THB per transaction, it is different from other figures.
- Finally, the overall figure, shown in Fig. 5(f), is comparable to Fig. 4(b). With 25.6% of the transactions linked to loss values, KB*** is rated top, followed by BB*, SC*, KK*, and KT* in that order, with respective rankings of 20.5%, 12.3%, 10.5%, and 10.1%.

*4.3  Statistics Based on Numbers of Account*:

The following details are derived from the results linked to the account numbers shown in Fig. 6:
- According to Fig. 6(a), KB*** is rated first with 29.0% of the bank accounts linked to Category A, Category A, or loss values between 100 and 1,000 THB each transaction, while SC*, BB*, KT*, and GS* are ranked second through fifth, with 15.7%, 15.2%, 13.0%, and 8.6%, respectively.
- When bank accounts associated with Category B or loss values between 1,001 and 10,000 THB per transaction were considered, it can be seen in Fig. 5(b) that KB*** is ranked first with 25,5% of the accounts associated with the loss values, higher than BB*, which is in second place with 18,3%. SC*, KK*, and KT* are in third through fifth place, respectively, with 13,1%, 11,4%, and 9,2% of the loss values.
- However, KB*** is in first place with 25.8% of the accounts associated with the loss values for the transactions associated with Category B, or the loss values between 10,001 and 100,000 THB, as shown in Fig. 5(c). This is the same percentage as shown in Fig. 6(c), the transactions of KB** associated with the loss values of 1,001 to 10,000 THB. While BB*, KK*, SC*, and KT* are in second through sixth place, respectively, with 20.8%, 12.9%, 12.4%, and 8.9%.
- It is comparable to Fig. 6(d) for Fig. 6(d). With 28.1% of the accounts linked with Category B, or the loss values between 100,001 and 1,000,000 THB per transaction, KB*** is rated first. BB*, KK*, SC*, and KT* are listed from second to fifth with 19.0%, 13.8%, 12.0%, and 9.0% of the loss values, respectively.
- For Category E, the loss values between 100,000 and 1,000,000 THB per transaction, it is similar to Fig. 6(a) and Fig. 5(e), and Fig. 6(e) shows that KB*** is still in first place with 31.0% of the loss values, but SC*, BB*, KT*, and GS* are surprisingly in second to fifth place with 17.1%, 14.9%, 11.2%, and 7.5% of the loss values, respectively.
- Finally, Fig. 6(e) is comparable to Fig. 4(f) overall. BB*, SC*, KT*, and GS* are rated as the second to fifth position with 16.0%, 14.8%, 11.9%, and 7.6% of the transactions connected with the loss values, respectively. KB*** is placed first with 28.2% of the transactions related with the loss values.

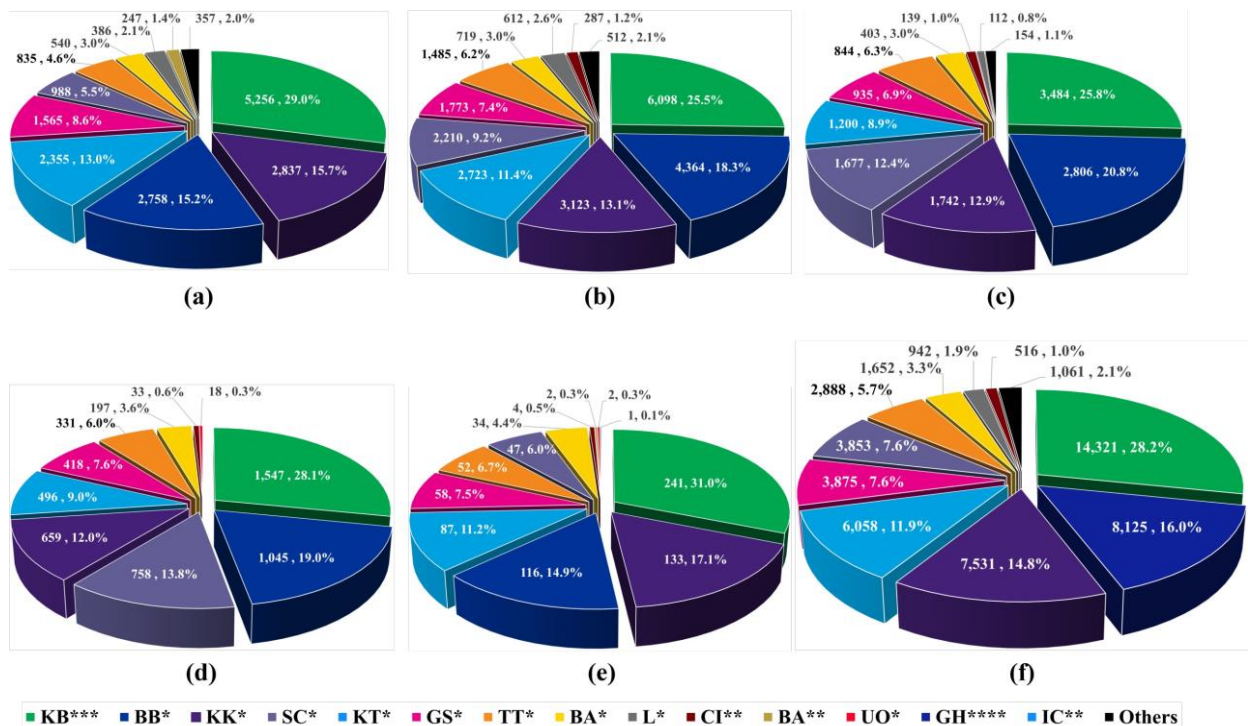

Fig.6. The statistics based on numbers of accounts (THB per transactions) by bank (a) 100-1,000 (b) 1,001-10,000 (c) 10,001-100,000 (d) 100,001-1,000,000 (e) 1,000,001-10,000,000 (f) Overall

*4.4 Comparison Among Five Categories of Loss Value per Transactions*

According to Fig. 4-6, the ranges of loss value per transaction were divided into five groups: A (100–1,000 THB per transaction), B (1,001–10,000 THB per transaction), C (10,001–100,000 THB per transaction),…, and E (1,000,001–10,000,000 THB per transaction). As a result, in this subsection, those five categories—which are related to lost values, transactions, and bank accounts—were taken into account and presented as follows:

- According to loss values, as shown in Fig. 7(a), Category D (100,001-1,000,000 THB per transaction) occupies the largest portion of the graph with 49.3% (2,308,400,224 THB), followed by Category E in second place with 32.4% (more than 1.517 billion THB), Category C in third place with 15.2% (more than 713 million THB), and Category B in fourth place with 2.8% (more than 132 million THB). The smallest section, Category A, only makes up 0.3% of the chart (or 12.8 million THB).
- In Fig. 7(b), the largest category by number of transactions is Category B, which accounts for 39.0% (34,304 transactions). Category A and C are in second and third place, respectively, with 28.5% (25,078 transactions) and 22.3% (19,591 transactions). While Category D and E make up the fourth and final portions with respective percentages of 9.0% (7,877 transactions) and 1.2% (1,093 transactions).
- In accordance with bank accounts, Fig. 7(c) is comparable to Fig. 7(b), Category B occupies the largest percentage, accounting for 38.7% (23,906 accounts), while Category A and C come in second and third, accounting for 29.3% (18,124 accounts) and 21.8% (13,496 accounts), respectively. While the fourth and final portions are Category D transactions between 100,000 and 1,000,000 THB and Category E, with rates of 8.9% (5,502 accounts) and 1.3% (777 accounts), respectively.

## 5. Discussions

From the results presented in Section 4, there are several key points that can be discussed as follows:
- From Fig. 4 through Fig. 6, KB*** and BB* were ranked top and second overall, respectively, highlighting the reliability of the findings. These statistics also show that both banks are the most frequently used in online frauds and scams.
- KB*** and BB* accounted for roughly 28.2% + 16.0% = 44.2% of scammer accounts, which is more than 22,000 accounts, as well as 25.6% + 20.5% = 46.1% of all transactions, which is more than 40,500 transactions, and 35.7% + 14.9% = 50.6% of all losses, which is more than 2,900 million THB.
- For overall, SC* is the third in terms of loss values (13.2%), transactions (12.3%), and number of accounts (14.8%).
- For overall, KT* is the fourth place when considering the transactions (11.3%) and the number of accounts (11.9%), but ranked the place fifth when it comes to the number of transactions (10.1%).
- It's surprising that while KKP is placed third in Fig. 5(f) based on the volume of transactions (10.5%), it drops to sixth place in Figs. 4(f) and 6(f), which take loss values (6.9%) and account volume (7.6%) into account. However, despite GS* placing fifth overall in terms of loss value (8.4%) and fraudulent accounts (7.6%), it placed sixth when considering the volume of transactions (7.9%), as seen in Fig. 5(f).
- Category D (100,001–1,000,000 THB per transaction) and Category E (1,000,001–10,000,000 THB per transaction) occupy 49.3% + 32.4% = 81.7%, or more than four fifths of the chart, according to Fig. 7(a). That indicates that Category D has the greatest economic impact.
- Since Category B occupies 39.0% of Fig. 7(b) and 38.7% of Fig. 7(c), which are linked with transactions and accounts, respectively, one can observe that they are consistent with one another. Additionally, Category A and B cover 28.5% + 39.0% = 67.5% of the transactions in Fig. 7(b). If a victim transfers money just once, it might be inferred that Category A and B account for the majority of the victims.
- With more attention paid to Categories A and B in Figures 7(b) and 8(c), it can be seen that there are 1.41 transactions per account, or (25,078 + 34,304) / (18,124 + 23,906). This number is not huge, but if banks cut back on the number of accounts used by scammers or money launderers, the volume of transactions and loss amounts will also go down.

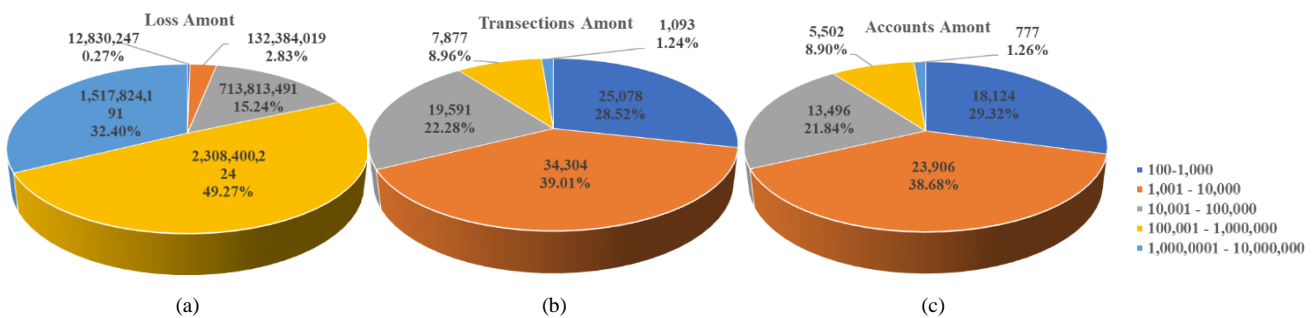

Fig. 7 Portion of each category of loss value per transaction and (a) to talal loss value (b) total transaction (c) total account

- In all, focused on KB*** in Figs. 4(f), 5(f), and 6(f), there were 22,552 total transactions, 14,321 total accounts, and a total loss value of 2,077,239,053 THB. The average is 145.048.46 THB per account or 92.108.86 THB each transaction. These are large numbers. This implies that financial victimizations are not recorded or complained about.
- Based on the data, the top six banks involved in online scams, KB***, BB*, SC*, KT*, GS*, and KK*, should collaborate closely with BOT, the Thai bank regulator, to develop new measures against money mules and the money laundering process (e.g., do not permit a new bank customer to enable a new online account without money as some banks do!). In order to curtail and halt illicit transactions, cyber frauds, and online scams in general, the government should also think about drafting and passing new legislation.
- It should be mentioned that only the data from ChaladOhn's database from February 2022 to January 2023 were the subject of this study's investigation and analysis. The loss values shown in this paper (ranging from 100 to 1,000,000 THB per transaction) cannot be taken to be an accurate representation of all loss values due to online fraud in Thailand. The percentage of money mule accounts in Thailand overall that are scam accounts is likewise unknown.
- The approach in this study might be applied to other countries if they have available database that is similar to ChaladOhn database. Therefore, the findings from that analysis can be utilized for new measures or regulations.

## 6. Conclusions

This study has investigated the ChaladOhn database, looking at 89,000 unlawful transactions in Thailand with average losses of less than 10 million THB. It has been determined that from February 2022 to January 2023, internet scam losses totaled more than 3,100 million THB. Furthermore, it was discovered that the KB*** bank was the most frequently used by scammers, accounting for 28.2%, 25.6%, and 35.7% of all bank accounts, transactions, and total loss value respectively, while the BB* bank came in the second place with 16.0%, 20.5%, and 14.9% of all accounts, transactions, and total loss value, respectively. Therefore, The Bank of Thailand (BOT) and/or the Thai Bankers' Association should consider this fact along with other findings in this work, then propose and enforce appropriate measures and regulations to prevent money laundering and reduce illegal transactions as a result of online scamming, especially KB*** and other financial institutions in the top lists. These measures and regulations should aim to prevent money laundering and reduce online scamming. However, only the data from the ChaladOhn system are taken into account in this analysis. In the future, data from other trustworthy sources should also be taken into account, followed by in-depth research and investigation to enable regulators better understand the behavior of online scams, and finally, the proposal and enforcement of new effective measures against this category of organized crime.